\documentclass[a4paper]{article}

\usepackage{INTERSPEECH2021}
\usepackage{multirow}
\usepackage[export]{adjustbox}

\title{The DKU System Description for The Interspeech 2021 Auto-KWS Challenge}

\name{Yechen Wang, Yan Jia, Murong Ma, Zexin Cai, Ming Li}
\address{Data Science Research Center, Duke Kunshan University, Kunshan, China }
\email{ming.li369@dukekunshan.edu.cn}
\begin{document}

\maketitle
\begin{abstract}
This paper introduces the system submitted by the DKU-SMIIP team for the Auto-KWS 2021 Challenge. Our implementation consists of a two-stage keyword spotting system based on query-by-example spoken term detection and a speaker verification system. We employ two different detection algorithms in our proposed keyword spotting system. The first stage adopts subsequence dynamic time warping for template matching based on frame-level language-independent bottleneck feature and phoneme posterior probability. We use a sliding window template matching algorithm based on acoustic word embeddings to further verify the detection from the first stage. As a result, our KWS system achieves an average score of 0.61 on the feedback dataset, which outperforms the baseline1 system by 0.25.
   
\end{abstract}
\noindent\textbf{Index Terms}: Keyword Spotting, Query-by-example Spoken Term Detection, Speaker Verification, Acoustic Word Embedding

\section{Introduction}

Keyword spotting (KWS), in terms of speech-level, is a task that detects whether a predefined word or phrase has appeared in continuous speech. It is commonly used as the primary technique for low resource trigger systems and speech-based document analysis. More recently, KWS has been widely applied to our daily life, such as the wake-up word detection module for speech assistants on mobile phones, vehicles, and smart speakers. Those speech assistants are triggered by predefined keywords, like "Hey, Cortana," "Alexa," and "Hey, Siri," spoken by the owner. Such applications raise the need for a customized KWS system that could detect the keyword and identify the target speaker's voice simultaneously. To this end, more attention has been paid to develop a KWS system that responds to a particular speaker in recent research \cite{12}. 

Conventionally, the KWS system consists of a large vocabulary continuous speech recognition (LVCSR) module, followed by a keyword spotting module that searches for keywords in the lattice generated by the LVSCR module \cite{chen2013quantifying}. The LVCSR module uses a large amount of audio-text pairs to train a traditional automatic speech recognition model that generates lattice, which contains the decoding information of the given speech. The following KWS module makes index for the lattice in the lattice and searches keywords accordingly. This provides a high accuracy approach while allowing us to customize the keywords without having to retrain the model. However, the lattice generated by the LVCSR module could be very complex with redundant information, which might result in inefficiency in real applications. Also this approach has relatively low performance when it comes to out-of-vocabulary (OOV) words because its highly dependency on the LVCSR module. 

Another widely used approach for KWS is the query-by-example spoken term detection (QbE-STD) system. The QbE-STD method detects keywords through efficient template matching based on linguistic features extracted from the speech. Typically, the QbE-STD method contains two steps, known as feature extraction and template matching. For the first step, the feature that represents the content of the reference keyword audio segment is extracted. Various features have been investigated as the representation in the literature. For example, supervised features like language-independent bottleneck feature (BNF) and phoneme posterior probability (PPP) is adopted and yielded relatively good performance \cite{1}. So as unsupervised features such as Mel-frequency cepstrum coefficients (MFCC)) \cite{3} and acoustic word embedding (AWE) \cite{4,5,6,7,8}. The keyword detection is completed by the second step using template matching algorithms. Given a pair of features extracted respectively from the template speech and the evaluated speech, matching algorithms such as segmental dynamic time warping \cite{9} and subsequence dynamic time warping (SDTW) \cite{9,10,11} are applied to measure the content similarity of the pair. The detection result is made according to the similarity score obtained from the matching algorithms. In the Auto-KWS 2021 Challenge, participants are required to develop a customized KWS approach that could efficiently handle the multilingual, multi accent and various keywords situation both in time and accuracy. In this case, we choose the QbE-STD approach to develop our entry for the challenge. 

We proposed a two-stage QbE-STD approach as our KWS system. Each stage contains a different QbE-STD system that consists of a feature extraction module and a template matching module. The first stage uses a BNF+PPP feature extractor in frame-level and an SDTW template matching algorithm. The second stage uses a sequence-level AWE feature extractor with a sliding window template matching algorithm. The search content must pass through both two stages. To achieve personalized keyword spotting, we employed the SV system that achieves state-of-the-art performance proposed in \cite{12} to determine whether the given speech and the template come from the same speaker. Our system obtains a final score of 0.61, with an average miss rate (MR) of 0.29 and an average false alarm rate (FAR) of 0.036 on the feedback dataset.

The paper is organized as follows. The detailed description of our submitted entry is presented in section 2. Section 3 describes the experimental setup, evaluation metrics, and our experimental results. Finally, the conclusion is provided in section 4. 

\section{System Description}

\subsection{KWS System}

A two-stage QbE-STD structure is adopted for our KWS system. In each stage, we applied a separated QbE-STD system with different feature extractors and template matching algorithms. When the keyword is detected by the first-stage model, the speech segment containing the keyword is fed to the second-stage model for another check. The second stage uses a sequence-level model with higher accuracy, but the output keyword time stamps of the second stage model are not as accurate as the first stage. Hence when the second stage also confirms the detection of the keyword, the keyword segment detected by the first-stage model is sent to the speaker verification module as its timing information is more accurate.



\subsubsection{First-stage Model} 
In general, two modules: feature extraction and template matching, are used in the  QbE-STD based KWS system. In the first stage, we use an acoustic model to extract the BNF and PPP feature. The time delay neural network (TDNN) based acoustic model has usually been applied in automatic speech recognition (ASR) tasks and achieves state-of-the-art performance. Therefore, we employ an acoustic modeling method based on the TDNN trained with frame-level training criteria.

\begin{figure*}[ht]
    \begin{center}
        \includegraphics[width=0.7\textwidth]{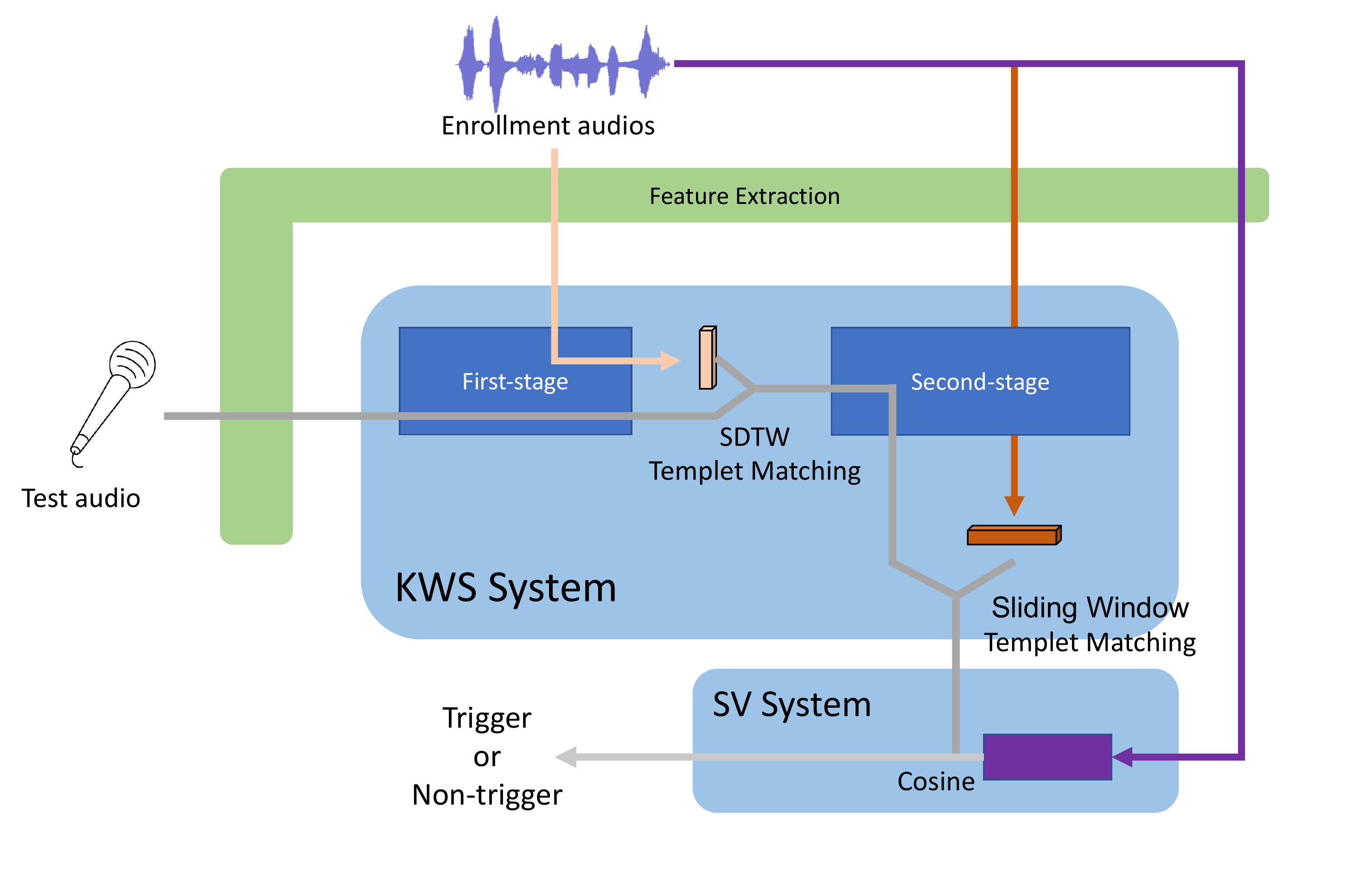}
    \end{center}
    \vspace*{-0.4cm}
    \caption{The overview framework of the proposed system.}
    \label{baseline_framework}
    \vspace*{-0.4cm}
\end{figure*}
\subsubsection{SDTW Template Matching Module} 

In the template matching step, a DTW-based multiple templates strategy is used in the first-stage KWS system\cite{chen2015query}. Since the QbE-STD tasks usually interfere with extraneous factors like channel variance, the templates fusion method has been widely used in the QbE-STD based system. Firstly, one of the prepared templates is chosen randomly as the master template, and then we apply the DTW algorithm to align the rest templates and get the shortest path. Finally, we compute the average of these aligned points in the shortest path and get the example template of the keyword. This fusion method allows us to obtain a more representative template though combining all the templates. 

Traditional DTW \cite{9} requires the start and the end time point of two sequences must be strictly aligned. In this challenge, we employ SDTW based algorithm \cite{10,11}. It can find a subsequence that does not necessarily go through the end time point, which most optimally fits the spoken query in the search content. The Euclidean distance is used in this SDTW template matching module. The alignment result is used as a keyword time stamp in future steps as the features we extracted in frame-level usually provide higher accuracy in the time axis during template matching.


\subsubsection{Second-stage Model} 
The second-stage keyword detection is activated after a successful trigger by the first stage. Similarly, we use two modules: an AWE system for feature extraction and a sliding window template matching method. Our AWE system uses a similar structure in \cite{8}. It is trained with sequence-level training criteria. The network consists of a combination of a convolutional neural network (CNN), a global average pooling layer (GAP), and a fully connected layer in order. The log filter-bank energies (Fbank) of individual words are extracted and fed into the network. The CNN structure is based on a residual neural network (ResNet) as it has been proved efficient in structuring deep neural networks. A global average pooling (GAP) layer is applied as an aggregator over the three-dimensional output sequence as it computes the global mean feature values over the time and frequency axes. The output of the GAP layer then goes through the fully-connected (FC) layer. We use the cross-entropy loss to optimize the system. The block softmax layer is also introduced in each ResNet block to better handle the multilingual scenario \cite{1,languageindependentbnf}. After the system is well-trained, we obtain the acoustic embedding feature from the output of the GAP layer.

\subsubsection{Sliding Window Template Matching Module} 
We choose a sliding window method and cosine distance as our template matching scheme. First, we pad or clip the keyword audio to 0.8 seconds, and use the same value as the fixed window size to convert the search content into a segment sequence $y_1,y_2,...y_t $. Then each segment is fed into the AWE feature extractor we trained to generate a sequence of acoustic word embedding features $f(y)=(f(y_1),f(y_2),...,f(y_t))$. For each input, the cost is calculated between a segment sequence of the input $f(x)$ and template $f(y)$ following the equation: 

$$ Cost(x,y)=\min(1-\frac{f(x)\cdot f(y_i)}{\left\|f(x)\right\|_2 \cdot \left\|f(y_i)\right\|_2}), i=1,2,...,T $$

It generates a score sequence with respect to time. The influence of random noise is removed by applying a simple moving average which could smooth the sequence by dividing the sum of a fixed number of continuous scores by the number of frames for the time involves. In addition, a template fusion method is also used in this module in order to find a more representative template. Since in this stage, the dimension of the feature extracted by the AWE system is a fixed value, we simply take the average of the templates as the fusion template.

\subsection{Speaker Verification System}

We use a similar model structure as it is proposed in \cite{14}. Specifically, our speaker verification system consists of a front-end feature extractor, a statistic pooling layer, and a back-end classifier. ResNet34 \cite{15} with SE-block \cite{16} is used as the feature extractor. An attentive statistics pooling (ASP) \cite{17} is used as the encoding layer, which has been proved efficient in detecting long-term speaker feature variations. For the back-end classifier, we use the AM-Softmax \cite{18}.

\subsection{Speaker Dependent KWS system}

Our proposed system consists of a two-stage KWS system and a speaker verification system described above. To achieve personalized KWS, as shown in Figure \ref{baseline_framework}, we design a speaker-dependent KWS system that only response to the target speaker when it detects the keyword. During the enrollment stage, we extract three utterances' embedding by the target speaker and save them as the enrollment speaker embedding vector template.


The general procedure of the system is as follows: First, in the two-stage KWS system, we save the timing information generate by the first stage frame-level SDTW alignment result. We use this information to cut a fixed-length segment of the speech features vector and fed it into the speaker verification system. The SV system then uses this vector to obtain the speaker embedding by feeding this acoustic vector into the SV model we described above, and the cosine similarity between this speaker embedding and the template we saved during the enrollment phase is compared with a threshold to determine if the input speech comes from the enrolled speaker. Together with the result of the two-stage KWS system and the result of the SV system, we consider an input speech to be positive if and only if it can successfully pass both the KWS system and the SV system. The general system diagram is shown in Figure \ref{baseline_framework}.

\section{Experimental Setup and Results}

\subsection{Experimental Setup}

We separated the training process of the two-stage KWS system and the SV system. Two systems are optimized separately. The training dataset provided by the Auto-KWS challenge organizer contains speech from 100 speakers recorded by mobile phones at a near-field around 0.2 meters. The audio has a single-channel 16-bit stream, and the sample rate is 16kHz. For each speaker, there are 10 enrollment utterances which contain the keyword, and a few others utterance that does not contain the keyword. Data augmentation is applied during the experiment to obtain more training data and improve the model accuracy and robustness. The data augmentation methods include perturbing the speed and the volume of the speech, adding noise, and splicing processing for the short speech audios during the enrollment. The practice dataset, which contains speech audio from 5 speakers, is also used in the evaluation.

\subsubsection{KWS System}
For our two-stage KWS system, in the first stage, we trained a frame-level TDNN based acoustic model on 40-dim MFCC features with a 25ms window length and a shift of 10ms. The training dataset includes multiple Chinese corpora on OpenSLR including Aidatatang \cite{aidatatang}, Aishell \cite{aishell_2017}, MagicData \cite{magicdata}, Primewords \cite{19}, ST-CMDS \cite{STCMDS}  and THCHS-30 \cite{THCHS30_2015}. The datasets we have used for training the acoustic model are shown in Table \ref{kws_dataset}. We trained a 3-gram language model using all the training transcriptions we have in the dataset. The lexicon is the CC-CEDIT Chinese dictionary expanded by Grapheme-to-Phoneme (G2P). The training starts by using a small part of data to accelerate the training procedure of the GMM model and then employ speaker adaptive training using all the datasets listed above. Finally, a Chain model is trained and evaluated while the PPP and BNF features are extracted from the final Chain model in the way of online decoding using Kaldi scripts \cite{Povey_ASRU2011}. The BNF and PPP features are stacked together and be used for the SDTW algorithm computation. A threshold is used here to decide if the input speech contains the keyword. We also save the time stamps on the shortest path for the later SV system as its timing information is more accurate.
\begin{table}[h]
    \centering
    \caption{The data used for training the acoustic model}        
    \label{kws_dataset}
    \begin{tabular}{cccccc}
        \toprule
        Dataset & Total hours\\
        \midrule
        Aidatatang & 140  \\
        Aishell & 151  \\
        MagicData& 712 \\
        Primewords & 99 \\
        ST-CMDS & 110  \\
        THCHS-30 & 26  \\
        \bottomrule
    \end{tabular}
    \vspace*{-0.4cm}
\end{table}
\begin{table*}[h]
    \centering
    \caption{Detailed result on practice and feedback dataset }        
    \label{result}
    \begin{tabular}{cccccc}
        \toprule
		Model & Dataset & Average Score & Average MR &	Average FAR \\
		\midrule
		\multirow{2}*{Our System}  &  Practice Dataset & 0.240  & 0.240  & 0 \\ & Feedback Dataset & 0.611	&0.289	&0.0359\\ 
		\midrule
		\multirow{2}*{Baseline System 1}  &  Practice Dataset & - & - & - \\ & Feedback Dataset & 0.859	&0.443	& 0.046\\ 
		\midrule
		\multirow{2}*{Baseline System 2}  &  Practice Dataset & - & - & - \\ & Feedback Dataset & 1.695	&0.481	& 0.135\\ 
		
        \bottomrule
    \end{tabular}
    \vspace*{-0.4cm}
\end{table*}

In the second stage, we use the 64-dimensional Fbank energies as the input acoustic feature for the AWE model. We use 0.8 seconds as the window length and extract acoustic features on this window. The proposed neural network architecture is shown in Table \ref{AWE} as L stands for input size.

\begin{table}[h]
\centering
\caption{The architecture for AWE system} 
\label{AWE}
\setlength{\tabcolsep}{1.5mm}{
\begin{tabular}{|c|c|c|c|c|}
\hline 
Layer&Output Size&Downsample&Channels&Blocks \\
\hline 
Conv1&$16\times \frac{L}{4}$&False&64&-\\
\hline 
Res1&$16\times \frac{L}{4}$&False&64&3\\
\hline 
Res2&$8\times \frac{L}{8}$&True&128&4\\
\hline  
Res3&$4\times \frac{L}{16}$&True&256&6\\
\hline 
Res3&$2\times \frac{L}{32}$&True&512&3\\
\hline 
GAP&512&-&-&-\\
\hline 
Output&Number of Words&-&-&-\\
\hline 
\end{tabular}}
\end{table}

The whole neural network was trained for 80 epochs with categorical cross-entropy loss optimized by Stochastic Gradient Descent (SGD) with Nesterov momentum equal to 0.9. We set the learning rate equal to 0.1 at first and gradually decrease its value every time the loss stops decreasing. The procedure of the sliding window template matching has been introduced above. The second threshold is used in the second stage to decide if the input speech contains the keyword.

\begin{table}[h]
    \centering
    \caption{The data used for pre-training}        
    \label{SV-pretrain-dataset}
    \begin{tabular}{cccccc}
        \toprule
          & Number of Speakers & Total hours & Utterances\\
        \midrule
        SLR38 & 855 & 100+ & 102600 \\
        SLR47 & 296 & 100+ & 50384 \\
        SLR62 & 600 & 200 & 237265 \\
        SLR62 & 274 & 1000 & 130108  \\
        SLR85 & 340 & 1500 & 108678 \\
        \bottomrule
    \end{tabular}
    \vspace*{-0.4cm}
\end{table}

\subsubsection{Speaker Verification}
The general experimental of our SV system has the same procedure as in \cite{12}. We pre-train our model by using data from SLR38 \cite{STCMDS}, SLR47 \cite{19}, SLR62 \cite{aidatatang}, SLR82 \cite{20}, SLR85 \cite{21} on OpenSLR. The datasets we have used for pre-training the SV system are shown in Table \ref{SV-pretrain-dataset}. We also add MUSAN \cite{musan2015} and RIRs-NOISES \cite{ko2017study} as noise in training set to make our model more robust. We set the signal-to-noise-ratio (SNR) between 0 to 20 dB while pre-training and 0 to 15 dB while fine-tuning. The method in \cite{14} is also applied to add reverberation to the data. During the pre-training process, we trained our model for 50 epochs also with an SGD optimizer together with a batch size of 256 and set initial learning rate equal to 0.01 and decreases 0.1 after every 20 epochs. We fine-tuned our model for 20 epochs with a learning rate of 0.001. The third threshold is used here to decide if the input speech comes from the target speaker.

\subsection{Evaluation Metrics and Determination of Thresholds}
In this challenge, the metrics defined by the organizers calculate the score from a weighted sum of the MR and the FAR follows the equation below:
$$score_i=MR+\alpha \times FAR$$
where $\alpha$ is a factor used to adjust the cost of MR and FAR. The lower the $score_i$, the better we consider the model is. Since we can adjust our three thresholds to make the trade-off between the MR and FAR, we fine-tune our model to achieve lower $score_i$ and higher performance. We designed a development dataset in order to fine-tune our three thresholds. For the KWS system, for each target speaker in the Auto-KWS Challenge training set, we directly select 5 enrollment utterances that contain the keyword and 20 other utterances that do not contain the keyword and randomly splice them together to create a new development dataset. For the SV system, we generate around 2800 trials using Auto-KWS Challenge training set to determine the threshold according to EER (Equal Error Rate) and minDCF \cite{greenberg20132012} performance. We finally choose $\alpha = 9$ and use the development set to find the optimal thresholds which could minimize our $score_i$.

\subsection{Experimental Results}

\subsubsection{SV Results}

The performance of our SV system on development data is shown in Table \ref{result_sv_baseline}. The threshold of the speaker verification system was determined by the development set. The mean threshold of EER and minDCF\cite{greenberg20132012} denoted as $(threshold_{EER}+threshold_{minDCF})/2$ have been used in our system. 

\begin{table}[h]
    \footnotesize
    \centering
    \caption{Performances of speaker verification  model on the dev sets (EER$[\%]$ and minDCF)}
    \label{result_sv_baseline}
    \begin{tabular}{cccc}
        \toprule
        Model & EER & minDCF  \\
        \midrule
        SV System  & 4.85 & 0.39 \\
        \bottomrule
    \end{tabular}
    \vspace*{-0.4cm}
\end{table}

\subsubsection{Results of the Overall System}
Our proposed system on the feedback dataset achieves an average score of 0.611. The detailed result is shown in Table \ref{result}. From the table, we can obtain the following observations from our system. First, using a more complex structure in the KWS system can achieve better results than the baseline systems. Our system achieves better performance than the baseline systems because we adopt the complex system structure of two-stage KWS models and large-scale speaker verification models. Second, only using the original training data to train our model, it is hard to generalize the model to the development and evaluation sets, which results in a very low recall. Thus, the method to determine the threshold is an important factor affecting the final score. The ad-hoc average threshold of EER and minDCF can improve system performance.

\section{Conclusions}
In this paper, we introduced the system we submitted to Auto-KWS 2021. Our system consists of a two-stage KWS system and an SV system. Although two systems are optimized separately, we managed to find an efficient way to work together and achieve personalized KWS. For the two-stage KWS system, we employed a BNF and PPP feature extractor together with an SDTW template matching method as the first stage and an AWE feature extractor with a sliding window template matching method as the second stage. Different fusion methods are applied to produce templates in different stages to improve performance. We also introduced data augmentation for the purpose of improving the accuracy and robustness of the system and designed the development dataset to optimize our score. Our system achieves an average score of 0.61 on the feedback dataset.


\bibliographystyle{IEEEtran}
\bibliography{template}

\begin{thebibliography}{10}
\providecommand{\url}[1]{#1}
\csname url@samestyle\endcsname
\providecommand{\newblock}{\relax}
\providecommand{\bibinfo}[2]{#2}
\providecommand{\BIBentrySTDinterwordspacing}{\spaceskip=0pt\relax}
\providecommand{\BIBentryALTinterwordstretchfactor}{4}
\providecommand{\BIBentryALTinterwordspacing}{\spaceskip=\fontdimen2\font plus
\BIBentryALTinterwordstretchfactor\fontdimen3\font minus
  \fontdimen4\font\relax}
\providecommand{\BIBforeignlanguage}[2]{{%
\expandafter\ifx\csname l@#1\endcsname\relax
\typeout{** WARNING: IEEEtran.bst: No hyphenation pattern has been}%
\typeout{** loaded for the language `#1'. Using the pattern for}%
\typeout{** the default language instead.}%
\else
\language=\csname l@#1\endcsname
\fi
#2}}
\providecommand{\BIBdecl}{\relax}
\BIBdecl

\bibitem{12}
Y.~Jia, X.~Wang, X.~Qin, Y.~Zhang, X.~Wang, J.~Wang, and M.~Li, ``The 2020
  personalized voice trigger challenge: Open database, evaluation metrics and
  the baseline systems,'' \emph{arXiv preprint arXiv:2101.01935}, 2021.

\bibitem{chen2013quantifying}
G.~Chen, S.~Khudanpur, D.~Povey, J.~Trmal, D.~Yarowsky, and O.~Yilmaz,
  ``Quantifying the value of pronunciation lexicons for keyword search in
  lowresource languages,'' in \emph{2013 IEEE International Conference on
  Acoustics, Speech and Signal Processing}.\hskip 1em plus 0.5em minus
  0.4em\relax IEEE, 2013, pp. 8560--8564.

\bibitem{1}
J.~Hou, C.-C.~L. Van Tung~Pham, C.-C. Leung, L.~Wang, H.~Xu, H.~Lv, L.~Xie,
  Z.~Fu, C.~Ni, X.~Xiao \emph{et~al.}, ``The nni query-by-example system for
  mediaeval 2015.'' in \emph{MediaEval}, 2015.

\bibitem{3}
P.~Yang, C.-C. Leung, L.~Xie, B.~Ma, and H.~Li, ``Intrinsic spectral analysis
  based on temporal context features for query-by-example spoken term
  detection,'' in \emph{Fifteenth Annual Conference of the International Speech
  Communication Association}, 2014.

\bibitem{4}
H.~Kamper, W.~Wang, and K.~Livescu, ``Deep convolutional acoustic word
  embeddings using word-pair side information,'' in \emph{2016 IEEE
  International Conference on Acoustics, Speech and Signal Processing
  (ICASSP)}.\hskip 1em plus 0.5em minus 0.4em\relax IEEE, 2016, pp. 4950--4954.

\bibitem{5}
S.~Settle and K.~Livescu, ``Discriminative acoustic word embeddings: Tecurrent
  neural network-based approaches,'' in \emph{2016 IEEE Spoken Language
  Technology Workshop (SLT)}.\hskip 1em plus 0.5em minus 0.4em\relax IEEE,
  2016, pp. 503--510.

\bibitem{6}
S.~Settle, K.~Levin, H.~Kamper, and K.~Livescu, ``Query-by-example search with
  discriminative neural acoustic word embeddings,'' \emph{arXiv preprint
  arXiv:1706.03818}, 2017.

\bibitem{7}
K.~Levin, A.~Jansen, and B.~Van~Durme, ``Segmental acoustic indexing for zero
  resource keyword search,'' in \emph{2015 IEEE International Conference on
  Acoustics, Speech and Signal Processing (ICASSP)}.\hskip 1em plus 0.5em minus
  0.4em\relax IEEE, 2015, pp. 5828--5832.

\bibitem{8}
M.~{Ma}, H.~{Wu}, X.~{Wang}, L.~{Yang}, J.~{Wang}, and M.~{Li}, ``Acoustic word
  embedding system for code-switching query-by-example spoken term detection,''
  in \emph{2021 12th International Symposium on Chinese Spoken Language
  Processing (ISCSLP)}, 2021, pp. 1--5.

\bibitem{9}
M.~M{\"u}ller, ``Dynamic time warping,'' \emph{Information retrieval for music
  and motion}, pp. 69--84, 2007.

\bibitem{10}
X.~Anguera and M.~Ferrarons, ``Memory efficient subsequence dtw for
  query-by-example spoken term detection,'' in \emph{2013 IEEE International
  Conference on Multimedia and Expo (ICME)}.\hskip 1em plus 0.5em minus
  0.4em\relax IEEE, 2013, pp. 1--6.

\bibitem{11}
H.~Wu, M.~Li, Z.~Cai, and H.~Zhong, ``Unsupervised query by example spoken term
  detection using features concatenated with self-organizing map distances,''
  in \emph{2018 11th International Symposium on Chinese Spoken Language
  Processing (ISCSLP)}.\hskip 1em plus 0.5em minus 0.4em\relax IEEE, 2018, pp.
  1--5.

\bibitem{chen2015query}
G.~Chen, C.~Parada, and T.~N. Sainath, ``Query-by-example keyword spotting
  using long short-term memory networks,'' in \emph{2015 IEEE International
  Conference on Acoustics, Speech and Signal Processing (ICASSP)}.\hskip 1em
  plus 0.5em minus 0.4em\relax IEEE, 2015, pp. 5236--5240.

\bibitem{languageindependentbnf}
K.~Vesel{\`y}, M.~Karafi{\'a}t, F.~Gr{\'e}zl, M.~Janda, and E.~Egorova, ``The
  language-independent bottleneck features,'' in \emph{2012 IEEE Spoken
  Language Technology Workshop (SLT)}.\hskip 1em plus 0.5em minus 0.4em\relax
  IEEE, 2012, pp. 336--341.

\bibitem{14}
J.~S. Chung, J.~Huh, S.~Mun, M.~Lee, H.~S. Heo, S.~Choe, C.~Ham, S.~Jung, B.-J.
  Lee, and I.~Han, ``In defence of metric learning for speaker recognition,''
  \emph{arXiv preprint arXiv:2003.11982}, 2020.

\bibitem{15}
K.~He, X.~Zhang, S.~Ren, and J.~Sun, ``Deep residual learning for image
  recognition,'' in \emph{Proceedings of the IEEE conference on computer vision
  and pattern recognition}, 2016, pp. 770--778.

\bibitem{16}
J.~Hu, L.~Shen, and G.~Sun, ``Squeeze-and-excitation networks,'' in
  \emph{Proceedings of the IEEE conference on computer vision and pattern
  recognition}, 2018, pp. 7132--7141.

\bibitem{17}
K.~Okabe, T.~Koshinaka, and K.~Shinoda, ``Attentive statistics pooling for deep
  speaker embedding,'' \emph{arXiv preprint arXiv:1803.10963}, 2018.

\bibitem{18}
H.~Wang, Y.~Wang, Z.~Zhou, X.~Ji, D.~Gong, J.~Zhou, Z.~Li, and W.~Liu,
  ``Cosface: Large margin cosine loss for deep face recognition,'' in
  \emph{Proceedings of the IEEE conference on computer vision and pattern
  recognition}, 2018, pp. 5265--5274.

\bibitem{aidatatang}
``aidatatang\textunderscore200zh, a free {C}hinese {M}andarin speech corpus by
  {B}eijing {D}ata{T}ang {T}echnology {C}o., {L}td ( www.datatang.com ),''
  \url{http://www.openslr.org/62/} Accessed April 3, 2021.

\bibitem{aishell_2017}
H.~Bu, J.~Du, X.~Na, B.~Wu, and H.~Zheng, ``Aishell-1: An open-source mandarin
  speech corpus and a speech recognition baseline,'' in \emph{2017 20th
  Conference of the Oriental Chapter of the International Coordinating
  Committee on Speech Databases and Speech I/O Systems and Assessment
  (O-COCOSDA)}.\hskip 1em plus 0.5em minus 0.4em\relax IEEE, 2017, pp. 1--5.

\bibitem{magicdata}
``{M}agic {D}ata {T}echnology {C}o., {L}td.,,'' 2019,
  \url{http://www.openslr.org/68/} Accessed April 3, 2021.

\bibitem{19}
L.~Primewords Information Technology~Co., ``Primewords chinese corpus set 1,''
  2018, \url{https://www.primewords.cn}.

\bibitem{STCMDS}
``{ST}\text{-}{CMDS}\text{-}20170001\textunderscore1, {Free ST Chinese Mandarin
  Corpus},'' \url{https://www.openslr.org/38/} Accessed April 3, 2021.

\bibitem{THCHS30_2015}
D.~Wang and X.~Zhang, ``Thchs-30: A free chinese speech corpus,'' \emph{arXiv
  preprint arXiv:1512.01882}, 2015.

\bibitem{Povey_ASRU2011}
D.~Povey, A.~Ghoshal, G.~Boulianne, L.~Burget, O.~Glembek, N.~Goel,
  M.~Hannemann, P.~Motlicek, Y.~Qian, P.~Schwarz, J.~Silovsky, G.~Stemmer, and
  K.~Vesely, ``The kaldi speech recognition toolkit,'' in \emph{IEEE 2011
  Workshop on Automatic Speech Recognition and Understanding}.\hskip 1em plus
  0.5em minus 0.4em\relax IEEE Signal Processing Society, 2011.

\bibitem{20}
Y.~Fan, J.~Kang, L.~Li, K.~Li, H.~Chen, S.~Cheng, P.~Zhang, Z.~Zhou, Y.~Cai,
  and D.~Wang, ``Cn-celeb: a challenging chinese speaker recognition dataset,''
  in \emph{ICASSP 2020-2020 IEEE International Conference on Acoustics, Speech
  and Signal Processing (ICASSP)}.\hskip 1em plus 0.5em minus 0.4em\relax IEEE,
  2020, pp. 7604--7608.

\bibitem{21}
X.~Qin, H.~Bu, and M.~Li, ``Hi-mia: A far-field text-dependent speaker
  verification database and the baselines,'' in \emph{ICASSP 2020-2020 IEEE
  International Conference on Acoustics, Speech and Signal Processing
  (ICASSP)}.\hskip 1em plus 0.5em minus 0.4em\relax IEEE, 2020, pp. 7609--7613.

\bibitem{musan2015}
D.~Snyder, G.~Chen, and D.~Povey, ``{MUSAN}: {A} {M}usic, {S}peech, and {N}oise
  {C}orpus,'' 2015, arXiv:1510.08484v1.

\bibitem{ko2017study}
T.~Ko, V.~Peddinti, D.~Povey, M.~L. Seltzer, and S.~Khudanpur, ``A study on
  data augmentation of reverberant speech for robust speech recognition,'' in
  \emph{2017 IEEE International Conference on Acoustics, Speech and Signal
  Processing (ICASSP)}.\hskip 1em plus 0.5em minus 0.4em\relax IEEE, 2017, pp.
  5220--5224.

\bibitem{greenberg20132012}
C.~S. Greenberg, V.~M. Stanford, A.~F. Martin, M.~Yadagiri, G.~R. Doddington,
  J.~J. Godfrey, and J.~Hernandez-Cordero, ``The 2012 nist speaker recognition
  evaluation.'' in \emph{INTERSPEECH}, 2013, pp. 1971--1975.

\end{thebibliography}

\end{document}